%Paper: hep-ph/9304305
%From: TZANI%EBUBECM1.BITNET@FRMOP11.CNUSC.FR
%Date: Tue, 27 Apr 93 19:07:45 BCN

\font\twelverm=cmr10  scaled 1200   \font\twelvei=cmmi10  scaled 1200
\font\twelvesy=cmsy10 scaled 1200   \font\twelveex=cmex10 scaled 1200
\font\twelvebf=cmbx10 scaled 1200   \font\twelvesl=cmsl10 scaled 1200
\font\twelvett=cmtt10 scaled 1200   \font\twelveit=cmti10 scaled 1200
\font\twelvesc=cmcsc10 scaled 1200  \font\twelvesf=cmss10 scaled 1200
\skewchar\twelvei='177   \skewchar\twelvesy='60

%  Define \...point macros to change fonts and spacings consistently

\def\twelvepoint{\normalbaselineskip=12.4pt plus 0.1pt minus 0.1pt
  \abovedisplayskip 12.4pt plus 3pt minus 9pt
  \belowdisplayskip 12.4pt plus 3pt minus 9pt
  \abovedisplayshortskip 0pt plus 3pt
  \belowdisplayshortskip 7.2pt plus 3pt minus 4pt
  \smallskipamount=3.6pt plus1.2pt minus1.2pt
  \medskipamount=7.2pt plus2.4pt minus2.4pt
  \bigskipamount=14.4pt plus4.8pt minus4.8pt
  \def\rm{\fam0\twelverm}          \def\it{\fam\itfam\twelveit}%
  \def\sl{\fam\slfam\twelvesl}     \def\bf{\fam\bffam\twelvebf}%
  \def\mit{\fam 1}                 \def\cal{\fam 2}%
  \def\sc{\twelvesc}               \def\tt{\twelvett}
  \def\sf{\twelvesf}
  \textfont0=\twelverm   \scriptfont0=\tenrm   \scriptscriptfont0=\sevenrm
  \textfont1=\twelvei    \scriptfont1=\teni    \scriptscriptfont1=\seveni
  \textfont2=\twelvesy   \scriptfont2=\tensy   \scriptscriptfont2=\sevensy
  \textfont3=\twelveex   \scriptfont3=\twelveex  \scriptscriptfont3=\twelveex
  \textfont\itfam=\twelveit
  \textfont\slfam=\twelvesl
  \textfont\bffam=\twelvebf \scriptfont\bffam=\tenbf
  \scriptscriptfont\bffam=\sevenbf
  \normalbaselines\rm}

%       tenpoint

%%
%%      Various internal macros
%%

\def\beginlinemode{\endmode
  \begingroup\parskip=0pt \obeylines\def\\{\par}\def\endmode{\par\endgroup}}
\def\beginparmode{\endmode
  \begingroup \def\endmode{\par\endgroup}}
\let\endmode=\par
{\obeylines\gdef\
{}}
\def\singlespace{\baselineskip=\normalbaselineskip}

\def\oneandahalfspace{\baselineskip=\normalbaselineskip
  \multiply\baselineskip by 3 \divide\baselineskip by 2}
\def\doublespace{\baselineskip=\normalbaselineskip \multiply\baselineskip by 2}

\newcount\firstpageno
\firstpageno=2
%% FOLLOWING LINE CANNOT BE BROKEN BEFORE 80 CHAR
\footline={\ifnum\pageno<\firstpageno{\hfil}\else{\hfil\twelverm\folio\hfil}\fi}
\def\toppageno{\global\footline={\hfil}\global\headline
  ={\ifnum\pageno<\firstpageno{\hfil}\else{\hfil\twelverm\folio\hfil}\fi}}
\let\rawfootnote=\footnote              % We must set the footnote style
\def\footnote#1#2{{\rm\singlespace\parindent=0pt\parskip=0pt
  \rawfootnote{#1}{#2\hfill\vrule height 0pt depth 6pt width 0pt}}}
\def\raggedcenter{\leftskip=4em plus 12em \rightskip=\leftskip
  \parindent=0pt \parfillskip=0pt \spaceskip=.3333em \xspaceskip=.5em
  \pretolerance=9999 \tolerance=9999
  \hyphenpenalty=9999 \exhyphenpenalty=9999 }
\def\dateline{\rightline{\ifcase\month\or
  January\or February\or March\or April\or May\or June\or
  July\or August\or September\or October\or November\or December\fi
  \space\number\year}}
\def\received{\vskip 3pt plus 0.2fill
 \centerline{\sl (Received\space\ifcase\month\or
  January\or February\or March\or April\or May\or June\or
  July\or August\or September\or October\or November\or December\fi
  \qquad, \number\year)}}

%%
%%      Page layout, margins, font and spacing (feel free to change)
%%

\hsize=6.5truein
\hoffset=0.0truein
\vsize=8.9truein
\voffset=0.0truein
\parskip=\medskipamount
\def\\{\cr}
\twelvepoint            % selects twelvepoint fonts (cf. \tenpoint)
\doublespace            % selects double spacing for main part of paper (cf.
                        %       \singlespace, \oneandahalfspace)
\overfullrule=0pt       % delete the nasty little black boxes for overfull box

%%
%%      This used to be timestamp.tex
%%

\newcount\timehour
\newcount\timeminute
\newcount\timehourminute
\def\daytime{\timehour=\time\divide\timehour by 60
  \timehourminute=\timehour\multiply\timehourminute by-60
  \timeminute=\time\advance\timeminute by \timehourminute
  \number\timehour:\ifnum\timeminute<10{0}\fi\number\timeminute}
\def\today{\number\day\space\ifcase\month\or Jan\or Feb\or Mar
  \or Apr\or May\or Jun\or Jul\or Aug\or Sep\or Oct\or
  Nov\or Dec\fi\space\number\year}

  %  "Draft", Timestamp

%%
%%      The user definitions for major parts of a paper (feel free to change)
%%

      % Preprint number at upper right of title page

\def\ubec#1{
 \rightline{\rm UB--ECM--PF--#1}}      % Preprint number at upper right

       % Preprint number at upper right of title page

\def\title                      %  Title on title page
  {\null\vskip 3pt plus 0.2fill
   \beginlinemode \doublespace \raggedcenter \bf}

\def\author                     %  Author(s) name(s)  on title page
  {\vskip 3pt plus 0.2fill \beginlinemode
   \doublespace \raggedcenter}

\def\affil                      % Affiliations (can intermix with \author)
  {\vskip 3pt plus 0.1fill \beginlinemode
   \oneandahalfspace \raggedcenter \it}

\def\abstract                   % Begin abstract
  {\vskip 3pt plus 0.3fill \beginparmode \narrower
   \oneandahalfspace {\it  Abstract}:\  }

\def\endtopmatter               % End title page, begin body of paper
  {\endpage                     %       This subsumes \body
   \body}

\def\body                       % Begin text body;  can be used to end
  {\beginparmode}               % \title, \author, \affil, \abstract,
                                % \reference, or \figurecaption modes

\def\head#1{                    % Head;  NOTE enclose the text in {}
  \goodbreak\vskip 0.4truein    %  e.g., \head{I. Introduction}
  {\immediate\write16{#1}
   \raggedcenter {\sc #1} \par }
   \nobreak\vskip 0truein\nobreak}

\def\beneathrel#1\under#2{\mathrel{\mathop{#2}\limits_{#1}}}

\def\refto#1{$^{#1}$}           % For references in text as superscript

\def\references                 % Begin references -- basic format is Phys Rev
  {\head{References}            % i.e., volume, page, year (space after
%%commas).
   \beginparmode
   \frenchspacing \parindent=0pt    %\leftskip=1truecm ?
   \parskip=0pt \everypar{\hangindent=20pt\hangafter=1}}

\gdef\refis#1{\item{#1.\ }}                     % Ref list numbers.

\gdef\journal#1,#2,#3,#4.{              % Journal reference.  Comma sets
    {\sl #1~}{\bf #2}, #3 (#4).}                % off: name, vol, page, year

\def\endreferences{\body}

\def\figurecaptions             % Begin figure captions
  {\endpage
   \beginparmode
   \head{Figure Captions}
}

\def\endpage                    %  Eject a page
  {\vfill\eject}

\def\endpaper                   %  Ways to say goodbye
  {\endmode\vfill\supereject}

%%
%%      Various little user definitions
%%

\def\ref#1{Ref.~#1}                     %       for inline references
\def\Ref#1{Ref.~#1}                     %       ditto
\def\[#1]{[\cite{#1}]}
\def\cite#1{{#1}}
          % For citation of equation numbers
        %       ditto
                     %       ditto
                   %       ditto
\def\(#1){(\call{#1})}
\def\call#1{{#1}}
\def\taghead#1{}
\def\frac#1#2{{#1 \over #2}}

\def\12{{1\over2}}

\def\sla{\raise.15ex\hbox{$/$}\kern-.57em}
\def\leaderfill{\leaders\hbox to 1em{\hss.\hss}\hfill}
\def\twiddle{\lower.9ex\rlap{$\kern-.1em\scriptstyle\sim$}}
\def\bigtwiddle{\lower1.ex\rlap{$\sim$}}
\def\gtwid{\mathrel{\raise.3ex\hbox{$>$\kern-.75em\lower1ex\hbox{$\sim$}}}}
\def\ltwid{\mathrel{\raise.3ex\hbox{$<$\kern-.75em\lower1ex\hbox{$\sim$}}}}
\def\square{\kern1pt\vbox{\hrule height 1.2pt\hbox{\vrule width 1.2pt\hskip 3pt
   \vbox{\vskip 6pt}\hskip 3pt\vrule width 0.6pt}\hrule height 0.6pt}\kern1pt}
\def\tdot#1{\mathord{\mathop{#1}\limits^{\kern2pt\ldots}}}

\def\pmb#1{\setbox0=\hbox{#1}%
  \kern-.025em\copy0\kern-\wd0
  \kern  .05em\copy0\kern-\wd0
  \kern-.025em\raise.0433em\box0 }

                           %%dalembertian, used to be \box

\def\crgrant{This research was supported in part by the CICYT
grant AEN90-0033. R.T. also acknowledges a grant from the
Ministerio de Educaci\'on y Ciencia, Spain}

\catcode`@=11
\newcount\tagnumber\tagnumber=0

\immediate\newwrite\eqnfile
\newif\if@qnfile\@qnfilefalse
\def\write@qn#1{}
\def\writenew@qn#1{}
\def\w@rnwrite#1{\write@qn{#1}\message{#1}}
\def\@rrwrite#1{\write@qn{#1}\errmessage{#1}}

\def\taghead#1{\gdef\t@ghead{#1}\global\tagnumber=0}
\def\t@ghead{}

\expandafter\def\csname @qnnum-3\endcsname
  {{\t@ghead\advance\tagnumber by -3\relax\number\tagnumber}}
\expandafter\def\csname @qnnum-2\endcsname
  {{\t@ghead\advance\tagnumber by -2\relax\number\tagnumber}}
\expandafter\def\csname @qnnum-1\endcsname
  {{\t@ghead\advance\tagnumber by -1\relax\number\tagnumber}}
\expandafter\def\csname @qnnum0\endcsname
  {\t@ghead\number\tagnumber}
\expandafter\def\csname @qnnum+1\endcsname
  {{\t@ghead\advance\tagnumber by 1\relax\number\tagnumber}}
\expandafter\def\csname @qnnum+2\endcsname
  {{\t@ghead\advance\tagnumber by 2\relax\number\tagnumber}}
\expandafter\def\csname @qnnum+3\endcsname
  {{\t@ghead\advance\tagnumber by 3\relax\number\tagnumber}}

\def\equationfile{%
  \@qnfiletrue\immediate\openout\eqnfile=\jobname.eqn%
  \def\write@qn##1{\if@qnfile\immediate\write\eqnfile{##1}\fi}
  \def\writenew@qn##1{\if@qnfile\immediate\write\eqnfile
    {\noexpand\tag{##1} = (\t@ghead\number\tagnumber)}\fi}
}

\def\callall#1{\xdef#1##1{#1{\noexpand\call{##1}}}}
\def\call#1{\each@rg\callr@nge{#1}}

\def\each@rg#1#2{{\let\thecsname=#1\expandafter\first@rg#2,\end,}}
\def\first@rg#1,{\thecsname{#1}\apply@rg}
\def\apply@rg#1,{\ifx\end#1\let\next=\relax%
\else,\thecsname{#1}\let\next=\apply@rg\fi\next}

\def\callr@nge#1{\calldor@nge#1-\end-}
\def\callr@ngeat#1\end-{#1}
\def\calldor@nge#1-#2-{\ifx\end#2\@qneatspace#1 %
  \else\calll@@p{#1}{#2}\callr@ngeat\fi}
\def\calll@@p#1#2{\ifnum#1>#2{\@rrwrite{Equation range #1-#2\space is bad.}
\errhelp{If you call a series of equations by the notation M-N, then M and
N must be integers, and N must be greater than or equal to M.}}\else%
 {\count0=#1\count1=#2\advance\count1
by1\relax\expandafter\@qncall\the\count0,%
  \loop\advance\count0 by1\relax%
    \ifnum\count0<\count1,\expandafter\@qncall\the\count0,%
  \repeat}\fi}

\def\@qneatspace#1#2 {\@qncall#1#2,}
\def\@qncall#1,{\ifunc@lled{#1}{\def\next{#1}\ifx\next\empty\else
  \w@rnwrite{Equation number \noexpand\(>>#1<<) has not been defined yet.}
  >>#1<<\fi}\else\csname @qnnum#1\endcsname\fi}

\let\eqnono=\eqno
\def\eqno(#1){\tag#1}
\def\tag#1$${\eqnono(\displayt@g#1 )$$}

\def\aligntag#1\endaligntag
  $${\gdef\tag##1\\{&(##1 )\cr}\eqalignno{#1\\}$$
  \gdef\tag##1$${\eqnono(\displayt@g##1 )$$}}

\def\eqalignno#1{\displ@y \tabskip\centering
  \halign to\displaywidth{\hfil$\displaystyle{##}$\tabskip\z@skip
    &$\displaystyle{{}##}$\hfil\tabskip\centering
    &\llap{$\displayt@gpar##$}\tabskip\z@skip\crcr
    #1\crcr}}

\def\displayt@gpar(#1){(\displayt@g#1 )}

\def\displayt@g#1 {\rm\ifunc@lled{#1}\global\advance\tagnumber by1
        {\def\next{#1}\ifx\next\empty\else\expandafter
        \xdef\csname @qnnum#1\endcsname{\t@ghead\number\tagnumber}\fi}%
  \writenew@qn{#1}\t@ghead\number\tagnumber\else
        {\edef\next{\t@ghead\number\tagnumber}%
        \expandafter\ifx\csname @qnnum#1\endcsname\next\else
        \w@rnwrite{Equation \noexpand\tag{#1} is a duplicate number.}\fi}%
  \csname @qnnum#1\endcsname\fi}

\def\ifunc@lled#1{\expandafter\ifx\csname @qnnum#1\endcsname\relax}

\let\@qnend=\end\gdef\end{\if@qnfile
\immediate\write16{Equation numbers written on []\jobname.EQN.}\fi\@qnend}

\catcode`@=12

\catcode`@=11
\newcount\r@fcount \r@fcount=0
\newcount\r@fcurr
\immediate\newwrite\reffile
\newif\ifr@ffile\r@ffilefalse
\def\w@rnwrite#1{\ifr@ffile\immediate\write\reffile{#1}\fi\message{#1}}

\def\writer@f#1>>{}
\def\referencefile{%                      Stuff to write .REF file
  \r@ffiletrue\immediate\openout\reffile=\jobname.ref%
  \def\writer@f##1>>{\ifr@ffile\immediate\write\reffile%
    {\noexpand\refis{##1} = \csname r@fnum##1\endcsname = %
     \expandafter\expandafter\expandafter\strip@t\expandafter%
     \meaning\csname r@ftext\csname r@fnum##1\endcsname\endcsname}\fi}%
  \def\strip@t##1>>{}}

\def\citeall#1{\xdef#1##1{#1{\noexpand\cite{##1}}}}
\def\cite#1{\each@rg\citer@nge{#1}}     % Variable No. of args, separated by
%%","

\def\each@rg#1#2{{\let\thecsname=#1\expandafter\first@rg#2,\end,}}
\def\first@rg#1,{\thecsname{#1}\apply@rg}       % each@ag is a general purpose
\def\apply@rg#1,{\ifx\end#1\let\next=\relax%      variable no. of arg. macro.
\else,\thecsname{#1}\let\next=\apply@rg\fi\next}% args separated by commas

\def\citer@nge#1{\citedor@nge#1-\end-}  % Check for M-N range (M and N numbers)
\def\citer@ngeat#1\end-{#1}
\def\citedor@nge#1-#2-{\ifx\end#2\r@featspace#1 % Single argument
  \else\citel@@p{#1}{#2}\citer@ngeat\fi}        % M-N range of arguments
\def\citel@@p#1#2{\ifnum#1>#2{\errmessage{Reference range #1-#2\space is bad.}
    \errhelp{If you cite a series of references by the notation M-N, then M and
    N must be integers, and N must be greater than or equal to
M.}}\else%
 {\count0=#1\count1=#2\advance\count1
by1\relax\expandafter\r@fcite\the\count0,%
  \loop\advance\count0 by1\relax%         Loop from M to N
    \ifnum\count0<\count1,\expandafter\r@fcite\the\count0,%
  \repeat}\fi}

\def\r@featspace#1#2 {\r@fcite#1#2,}    % Eat spaces at beginning or end of arg
\def\r@fcite#1,{\ifuncit@d{#1}          % Cite individual reference
    \expandafter\gdef\csname r@ftext\number\r@fcount\endcsname%
    {\message{Reference #1 to be supplied.}\writer@f#1>>#1 to be supplied.\par
     }\fi%
  \csname r@fnum#1\endcsname}

\def\ifuncit@d#1{\expandafter\ifx\csname r@fnum#1\endcsname\relax%
\global\advance\r@fcount by1%
\expandafter\xdef\csname r@fnum#1\endcsname{\number\r@fcount}}

\let\r@fis=\refis                       % Save old \refis, redefine
\def\refis#1#2#3\par{\ifuncit@d{#1}%      Use two params #2 #3 to strip blank
    \w@rnwrite{Reference #1=\number\r@fcount\space is not cited up to now.}\fi%
  \expandafter\gdef\csname r@ftext\csname r@fnum#1\endcsname\endcsname%
  {\writer@f#1>>#2#3\par}}

\def\r@ferr{\endreferences\errmessage{I was expecting to see
\noexpand\endreferences before now;  I have inserted it here.}}
\let\r@ferences=\references
\def\references{\r@ferences\def\endmode{\r@ferr\par\endgroup}}

\let\endr@ferences=\endreferences
\def\endreferences{\r@fcurr=0%            Save old \endreferences, redefine
  {\loop\ifnum\r@fcurr<\r@fcount%         Loop over refnum and produce text
    \advance\r@fcurr by 1\relax\expandafter\r@fis\expandafter{\number\r@fcurr}%
    \csname r@ftext\number\r@fcurr\endcsname%
  \repeat}\gdef\r@ferr{}\endr@ferences}

% Save old \endpaper, redefine it to write parting message.

\let\r@fend=\endpaper\gdef\endpaper{\ifr@ffile
\immediate\write16{Cross References written on []\jobname.REF.}\fi\r@fend}

\catcode`@=12

\citeall\refto          % These macros will generate citations
\citeall\ref            %
\citeall\Ref            %
%%
%%              TABLEOFC.TEX                    6/2/85  Doug Eardley
%%
%%              Used with JNL.TEX to produce a table of contents. To use,
%%      \input this file after JNL.  After running TEX <filename>,
%%      a new TeX file <filename>.TOC will be written, which contains the
%%      table of contents.  The latter file can then be edited (if necessary)
%%      and TEX'd itself.  All items #1 which appear in \head{#1} are listed.
%%              To list other items, use the macro \tocitem{#1}.
%%      To list automatically all items which appear in some macro \macroname,
%%      declare \tocitemall\macroname.  Ditto \tocitemitem and \tocitemitemall
%%      for subitems, and \tocitemitemitem and \tocitemitemitemall for subsub-
%%      items.  The macro \macroname must have one exactly one argument #1.
%%              At present, TABLEOFC is ***incompatible*** with PPT.

\catcode`@=11
\newwrite\tocfile\openout\tocfile=\jobname.toc
\newlinechar=`^^J
\write\tocfile{\string\input\space jnl^^J
  \string\pageno=-1\string\firstpageno=-1000\string\singlespace
  \string\null\string\vfill\string\centerline{TABLE OF CONTENTS}^^J
  \string\vskip 0.5 truein\string\rightline{\string\underbar{Page}}\smallskip}

\def\tocitem#1{%                Add item to table of contents, this page
  \t@cskip{#1}\bigskip}
\def\tocitemitem#1{%            Ditto subitem
  \t@cskip{\quad#1}\medskip}
\def\tocitemitemitem#1{%        Ditto subsubitem
  \t@cskip{\qquad#1}\smallskip}
\def\tocitemall#1{%             Make all macro#1 tocitem's
  \xdef#1##1{#1{##1}\noexpand\tocitem{##1}}}
\def\tocitemitemall#1{%         Make all macro#1 tocitemitem's
  \xdef#1##1{#1{##1}\noexpand\tocitemitem{##1}}}
\def\tocitemitemitemall#1{%             Make all macro#1 tocitemitemitem's
  \xdef#1##1{#1{##1}\noexpand\tocitemitemitem{##1}}}

\def\t@cskip#1#2{%              Add item to T.O.C with skip, this page
  \write\tocfile{\string#2\string\line{^^J
  #1\string\leaderfill\space\number\folio}}}

%\def\tocitempage#1#2{%         Add item to table of contents on page #2
%  \write\tocfile{\string\bigskip\string\line{^^J
%  #1\string\leaderfill\space\number#2}}}
%
%\newcount\pastpageno\pastpageno=0
%\def\tocitempast#1#2{%         Add item to toc, #2 pages past end of paper;
%  \ifnum\pastpageno=0
%    \global\pastpageno=\pageno
%    \global\advance\pastpageno by 1 \fi
%  \tocitempage{#1}\pastpageno \global\advance\pastpageno by #2}

\def\t@cproduce{%                 Spit out T.O.C.
  \write\tocfile{\string\vfill\string\vfill\string\supereject\string\end}
  \closeout\tocfile
  \immediate\write16{Table of Contents written on []\jobname.TOC.}}

% Save old \endpaper, redefine it to write .TOCfile.

\let\t@cend=\endpaper\def\endpaper{\t@cproduce\t@cend}

\catcode`@=12

\tocitemall\head                % \head{} will now generate T.O.C. entries

  \def\e{\epsilon}

\def \Dsl{\,\raise.15ex\hbox{$/$}\mkern-13.5mu {D}}
\def \DI{{\cal D}}
\def \dsl {\raise.15ex\hbox{$/$}\kern-.57em\hbox{$\partial$}}

\def \Del{\,\raise.15ex\hbox{$/$}\mkern-13.5mu {\Delta}}

\def \v {\rlap\slash{v}}

\def \el {\rlap\slash{\rm e}}

\ubec{93/1}
\title{Extra symmetries in the effective theory of heavy
quarks\footnote{$^*$}{\crgrant}}
\author{{\bf Joan Soto}\footnote\dag{soto@ebubecm1.bitnet} and
{\bf Rodanthy Tzani}\footnote\ddag{tzani@ebubecm1.bitnet}}
\affil{Departament d'Estructura i Constituents de la Mat\`eria,
Universitat de Barcelona, Barcelona, Catalonia, Spain}
\abstract{Extra symmetries are shown to exist in the effective theory
of heavy quarks when both quarks and anti-quarks with the same velocity
are included. These symmetries mix the quark with the anti-quark
sector and they resemble axial-type of symmetries. Together
with the known flavor and spin symmetries they form a $u(4)$ algebra
when a single flavor is considered. It is shown that the full $U(4)$
set of symmetries breaks spontaneously down to $ U(2) \otimes U(2)$.
The Goldstone modes corresponding to the spontaneously broken currents
are identified. Finally, the precise connection of this theory with
the fundamental QCD is derived and it is investigated under some
approximations. Some physical processes where these extra symmetries
may be relevant are pointed out.}

\endtopmatter

\body
\baselineskip=20pt
\null

{\bf 1. Introduction}
\bigskip
The physical properties of hadrons with a single heavy quark $Q$
($m_{Q}>>\Lambda_{QCD}$) are largely independent of the precise value of
the heavy quark mass $m_{Q}$ [1,2]. The heavy quark sector of the full
QCD lagrangian can be approximated by the so-called Heavy Quark Effective
Theory (HQET), in which the trivial leading dependence on the heavy
quark mass is removed [3-6] (see [7] for a review).
The most striking feature of the HQET lagrangian is that
it enjoys a number of symmetries which are absent in the original
QCD lagrangian. Moreover, finite mass corrections can be incorporated
systematically [8,9].

The flavor and spin symmetries have been known since
the early papers [2,5] and extensively used in phenomenological
applications [9-11]. These symmetries relate properties of hadrons
containing a single heavy quark of different spin and flavor,
which move with the same velocity with respect to a given reference
frame.

Since the HQET lagrangian has a natural separation in quark and
anti-quark sectors [5],
the above symmetries are also realized separately in
both sectors. Consequently, one can restrict oneself to either sector,
as it has been done in most of the applications studied by now.
This is physically reasonable since the HQET is a low
energy effective theory which is not able to account for heavy quark
pair production. (That would require an infinite momentum transfer in
the HQET framework.) While, however,
the lagrangian does not describe quark anti-quark pair production
it may well describe quarks and anti-quarks away from the
production point. Indeed, phenomenological applications of the HQET in
systems involving a heavy quark and a heavy
anti-quark, have recently been studied in the context of $D -\bar D$
and $B-\bar B$ mixing [12].

In the present work we carry out a theoretical study of the HQET with
both heavy quark and heavy anti-quark fields included in it. We are
concerned with the case of a single
flavor, although the case of $N_{f}$ flavors is briefly discussed in
the last section.

In the approach of HQET the heavy
quark field is characterized by a distinct fixed velocity and the
Hilbert space decomposes into superselection sectors labeled by the
velocity of the heavy quark.
In this paper we consider both quark and anti-quark in one of
these superselection sectors labeled by a fixed velocity $ v $.

In a previous work [13] we
showed that several symmetries, in particular the flavor and spin
symmetry, are anomaly free
in this approximation. In the same paper we pointed out that an extra
symmetry, which we called $\gamma _ 5 $-symmetry, exists in the HQET
when heavy quark and anti-quark with
the same velocity are included in it. This is an unexpected symmetry
because it mixes quark and anti-quark fields while particles and
antiparticles are not supposed to know about each other in the HQET.
In fact a larger set of symmetries of the same nature exist in this
theory in that case, that is, when quarks and anti-quarks are considered
with the same velocity. We devote this work to its discussion.

In section 2 we show  that the HQET describing quarks and
anti-quarks with the same velocity has a $U(4)$ symmetry if only one
flavor is considered. In section 3 we analyze the realization of this
symmetry and
show that it is spontaneously broken down to $U(2)\otimes U(2)$, i.e.,
to the flavor \footnote{$^{\dagger}$}{Since we restrict ourselves to a
single flavor, flavor symmetry is to be
understood as the $U(1)\otimes U(1)\subset U(2)\otimes U(2) $ symmetry
corresponding to the separate conservation of number particles and
antiparticles all over the paper.}
and spin symmetries. In section 4 we discuss the
implications of the Goldstone theorem and identify the Goldstone modes.
In section 5 we make the connection of this theory with the fundamental
QCD using the generating functional formalism. We conclude with a brief
discussion on
possible phenomenological applications of these symmetries in section 5.
                        \bigskip
{\bf 2. Extra symmetries in the HQET}

Consider the HQET describing  heavy quarks and anti-quarks with the
same velocity $v _ {\mu}$ ($ v _ {\mu} v ^{\mu} =1 $) [5]. It is given
by $$
L_v = i \bar h_v \v v _ {\mu} D ^ {\mu} h _ v \,= \, i \bar h ^+ _
v v \cdot D h ^+ _v \,- \, i \bar h^- _v v \cdot D h ^- _ v
\,,
\eqno(lag) $$
where $ h _ v = h ^+ _ v + h ^- _ v $ and $ h ^{\pm} _ v = { 1 {\pm} \v
\over 2 } h _ v $. $ h ^+ _ v $ contains creation operators of quarks
with small momentum about $ m v _ {\mu} $ and $ h ^- _v$ contains
annihilation operators of anti-quarks again with small momentum about $
m v _ {\mu} $. $D_{\mu}$ is the covariant derivative containing the
gluon field.

It is well-known that this theory is symmetric under rotations in the
spin and flavor space [2, 5, 7]. That is, the action is invariant under
the following transformations:
$$
h ^{\pm} _v \rightarrow e ^{i \epsilon ^i _ {\pm} S ^{\pm} _ i }
h ^{\pm} _v \; \; \;{\rm and } \;\;\; \bar h ^{\pm}_{v} \rightarrow \bar
h ^{\pm}_{v} e ^{ -i \e ^i _ {\pm} S ^{\pm} _ i }\; ,
\eqno(sp)$$
where $ S ^{\pm} _i = i \e _ {ijk} [ \el _ j , \el _k ] ( 1 \pm \v)
/2 \,$, with $ e ^{\mu} _ j \;, j = 1,2,3 $ being an orthonormal set of
space like vectors orthogonal to $ v _\mu \; $, and
$$
h ^{\pm} _v  \rightarrow e ^ {i \theta _{\pm}}  h ^{\pm}
_v  \; \;\; {\rm and } \;\;\; \bar h ^{\pm}_ v \rightarrow \bar h
^{\pm}_v  e ^{-i \theta _ {\pm}  }\; .
\eqno(fl)$$
 $ \epsilon ^i _ {\pm} $ and $
\theta _ {\pm} $ are arbitrary real numbers corresponding to
the parameters of the transformations.

It is important for what follows to emphasize that
the above symmetries are realized separately for the
quark and anti-quark sectors of the theory.
Namely, for every velocity $v$ there exists a $ U(2)
 $ symmetry for the quark sector of the lagrangian and a
$U(2)$ for the anti-quark sector, being the total
symmetry $ U(2) \otimes U (2)$. These symmetries are enlarged into
$U(2N_f) \otimes U (2N_f)$ when $N_f$ flavors are included in the
theory [5]. Note also that in terms of the field $h_v$ the last
transformations can be expressed as
$$
\eqalign{ h _ v &\rightarrow e ^{ i \theta } h _v  ; \;\;\;
\bar h _ v \rightarrow  \bar h _ v e ^{- i \theta }  \cr
h _ v &\rightarrow e ^{ i \v \theta } h _ v ; \;\;\;
\bar h _ v \rightarrow \bar h _v e ^{- i \v \theta } \cr }
\eqno(fl1)$$
for the flavor symmetry and
$$
\eqalign{ h _ v &\rightarrow e ^{ i \e ^i S _i } ; \;\;\;
\bar h _ v \rightarrow \bar h _ v e ^{- i \e ^i S _i }   \cr
h _ v &\rightarrow e ^{ i \e ^i S _ i \v } h _ v
; \;\;\;\bar h _ v \rightarrow \bar h _ v e ^{- i \e ^i S _ i \v } \cr}
\eqno(sp1)$$
for the spin symmetry, where now the spin operator is $ S_ i
= i {\e} _ {ijk} [ \el _ j , \el _ k ] $.
There are two generators $(i, i \v)$ for the flavor symmetry and six $(i
S _i, i S _i \v)$ for the spin symmetry. Note that the projection
operator $ (1 \pm \v)/2 $, which
was originally responsible for the separation of the lagrangian in $+$
and $-$ component terms, is hidden in the generators when
the symmetries are expressed in this basis.

Apart from the above symmetries the lagrangian
\(lag) is invariant under the following set of transformations:
$$
h _v \rightarrow e ^{i \gamma _ 5 \epsilon } h _ v \;\; \;\;{\rm ;}
\;\;\;\;
\bar h _ v \rightarrow \bar h _ v e ^{i \gamma _ 5 \epsilon } \, ,
\eqno(g5)$$
$$
h _ v \rightarrow e ^{ \gamma _ 5 \v \e } h _v \;\;\;\;{ \rm ;} \;\;
\;\; \bar h _v \rightarrow \bar h _v e ^{ \gamma _ 5 \v \e } \, ,
\eqno(gv)$$
$$
h _v \rightarrow e ^{ \e ^i \el _ i } h _ v \;\;\;\; {\rm ;}\;\; \;\;
\bar h _ v \rightarrow \bar h _ v e ^{ \e ^i \el _ i } \, ,
\eqno(ee)$$
$$
h _ v \rightarrow e ^{ i \e ^i \el _ i \v } h _ v \;\; \;\; {\rm ; }
\;\;\;\; \bar h _v \rightarrow \bar h _v e ^{i \e ^i \el _ i \v } \, .
\eqno(ev)$$

There are two observations to be made about these symmetries. Firstly,
they all mix quark and anti-quark sectors. Indeed, in terms of $h
^{\pm}_v$ fields these last transformations can be written as
$$
h ^{\pm} _ v \rightarrow cos \e \; h ^{\pm} _v \, + \, i \gamma _ 5 sin
\e \; h ^{\mp} _v \; ,
\eqno(g51)$$
$$
h ^{\pm} _v  \rightarrow cos \e \; h ^{\pm} _v \, + \, \gamma _ 5
\v sin \e \, h ^{\mp}_{v} \; ,
\eqno(gv1)$$
$$
h ^{\pm} _ v \rightarrow cos |\e| \; h ^{\pm} _ v \, + \, { \el _ i \e
^i \over |\e|} sin |\e| \; h ^{\mp}_{v} \; ,
\eqno(e1)$$
$$
h ^{\pm} _v \rightarrow cos |\e| \; h ^{\pm} + i {\el _i  \e ^ i \v
\over |\e|} sin |\e| \; h ^{\mp}_{v} \;
\eqno(ev1)$$
correspondingly, where $ |\e| := \sqrt {\e ^i \e ^i } $.
Secondly, they appear in sets, as in the case of the flavor \(fl) and spin
\(sp) symmetries. There are also eight generators for
this new set of symmetries, given by $(i \gamma _ 5 , \gamma _ 5 \v )$
and $( \el _i , i \el _i \v)$.

The fact that these symmetries transform quark fields into anti-quark
fields might suggest at first sight that they must be implemented
anti-unitarily at the level of the Hilbert space. However, since these
symmetries are {\it continuous} this possibility is ruled out.

In what follows we prove that the
transformations \(g5) - \(ev) together with \(fl1) - \(sp1)
can be accommodated in a $u(4)$ algebra. The explicit commutations
relations of all symmetry generators are given by
$$
[i, i \v] = [i, i \gamma _ 5] = [ i , \gamma _ 5 \v ] = [i ,
i S _i ] = [i, i S _ i \v] = [i , \el _ i ]= [i, i \el _ i \v] = 0 \;,
$$
$$
[i \v , i \gamma _ 5 ] = 2 \gamma _ 5 \v
,\; \;\; [i \v , \gamma _ 5 \v ] = -2i \gamma _ 5 ,\;\;\; [i \v , i S _i
] = 0,\;\;\; [i\v , i S _i \v ] = 0, \; \;\;
$$
$$
[i \v , \el _ i] = -2i \el _i \v,\;\;\;
[i \v , i \el _ i \v] = 2 \el _ i
$$
$$
[i \gamma _ 5 ,  \gamma _ 5 \v ] = 2i \v, \;\;\; [i \gamma _ 5 , i S
_ i ] = 0,\; \;\; [ i \gamma 5 , i S _ i \v ] = 8 \el _ i
$$
$$
[i \gamma _ 5 , \el _ i ] = - {i \over 2} S _ i \v \, ,\;\;\;
[i \gamma _ 5 , i \el _ i \v ] = 0
$$
$$
[\gamma _ 5 \v , i S _ i ] = 0 ,\;\;\; [ \gamma _ 5 \v , i S _ i \v ] =
-8i \el _ i \v , \;\; \;[ \gamma _ 5 \v , \el _ i ] = 0 , \; \; [ \gamma
_ 5 \v , i \el _ i \v ] = {i \over 2 } S _ i \v
$$
$$
[i S _ i , i S _ j ] = - 8 i \e ^{ijk} S _ k ,\;\; \;[i S _ i , i S _ j
\v] = -8i \e ^{ijk} S _ k \v , \;\;\; [i S _ i , \el _ j ] = -8 \e
^{ijk} \el _ k , \; \; \;
$$
$$
[i S _ i , i \el _ j \v ] = -8i \e ^{ijk } \el _ k \v
$$
$$
[i S _ i \v , i S _ j \v ] = - 8i \e ^{ijk} S _ k , \;\;\;
[i S _ i \v , \el _ j ] = 8i \delta _ {ij} \gamma _ 5 ,\;\;\;
[i S _ i \v , i \el _ j \v ] = -8 \delta _ {ij} \gamma _ 5 \v
$$
$$
[ \el _ i , \el _ j ] = -{i \over 2} \e ^{ijk} S _ k ,\;\;\;
[ \el _ i , i \el _ j \v ] = -2i \delta  _ {ij} \v
$$
$$
[i \el _ i \v , i \el _ j \v] = -{ i \over 2} \e ^{ijk} S _ k
\eqno(cev)$$

In order to prove that the last commutation relations obey the $u(4)$
algebra, it is more convenient to go to the rest frame. We choose the
following basis:
$$
v^ {\mu} = (1,0,0,0) ,\,\; \,e ^{\mu}_ 1 = (0,1,0,0), \;\,\, e^{\mu} _ 2
= (0,0,1,0) \;\,\,{\rm and }\,\,\; e ^{\mu}_ 3 = (0,0,0,1) \;.
\eqno(b)$$
Then the generators are reduced to the following set of $4 $x$ 4 $
matrices:
$$
 i, \; \;\; i \gamma _ 0 ,\; \;\; - \e ^{ijk } [ \gamma^{j} ,
\gamma^{k} ]
 ,\; \;\; - \e ^{ijk } [ \gamma^{j} , \gamma^{k} ]
\gamma_0=-4i\gamma_5\gamma^{i} ,\;
\;\; i \gamma _ 5 , \;\; \; \gamma _ 5 \gamma _ {0},  \;\;\; \gamma^{i},
\;\;\;
 i \gamma^{i} \gamma _ 0 . \;\;\;
\eqno(dm)$$
These are the 16 independent Dirac matrices, which are antihermitian
in the representation
where $\gamma _0 $ and $ \gamma _5 $ are hermitian, and the $
\gamma^{i} $ antihermitian. This set of matrices define
the $u(4)$ compact algebra. With this we conclude that our generators
satisfy the $ u(4)$ algebra in the rest frame. The fact that the algebra
remains the $u(4) $ in any other frame of reference follows from the
commutation relations. The structure constants are preserved in any
reference frame. This completes the proof.
(A more physical argument on the fact that our theory indeed has a
$u(4) $ algebra will be given later.)

Note at this point that not all transformations (6)-(9) are
unitary in an arbitrary reference
frame.
Only in the rest frame the generators of the $u(4)$ algebra have
definite hermiticity properties. This may seem to be in contradiction
with the fact that $u(4)$ is a
compact algebra. However, what
is guaranteed by general theorems is that finite dimensional
representations of compact Lie algebras (groups) are
equivalent to unitary representations, but not necessarily unitary
themselves. Therefore, the above generators in an arbitrary frame
constitute a non-unitary representation of $u(4)$, which should
certainly be equivalent to a unitary one.
(In fact, it is not difficul to explicitly
construct such a unitary representation.)
Moreover, it follows from (10)-(13) that $ e ^{ 2 \pi X } = 1$, for all
generators $X$ of the algebra. This proves that the group obtained by
exponenciating the algebra is compact, and hence it must be
identified with $U(4)$.

A straightforward consequence of these extra symmetries in this theory,
is the following:
In principle one could include a residual mass term of the form $
\delta m \bar h _ v
h _ v $ in the effective lagrangian since spin and flavor
symmetry allow it [14]. Such a mass term, however, would
break explicitly the extra symmetries that we pointed out above.
Enforcing these symmetries forbids such a residual mass term and, more
important, guarantees that it will not be induced by radiative
corrections.

Next we give the conserved currents for this new set of symmetries.
They are given by
$$
J ^{\mu} _ 5 = \bar h_{v} \v v ^{\mu} \gamma _5 h_{v} \;\;\; {\rm and
}\;\;\; J ^{\mu}_ {v5} = i \bar h_{v} ^{\mu} \gamma _5 h_{v}
\eqno (g5c)$$
and
$$
J ^{\mu} _ {5i} = i \bar h_{v} \v v ^{\mu} \el _ i h_{v} \;\;\; {\rm and
} \;\;\; J ^{\mu} _ {v5i} =\bar h_{v} v ^{\mu} \el _ i h_{v} \,
\eqno(ec)$$
for the $ \gamma _5 $ and $ \el _i $ sets of symmetries correspondingly.

It is not difficult to prove, following ref. [13], that none of these
currents has an anomaly. Notice that, due to the presence of the
operators $\gamma _5 $ and $ \el _ i$ which anticommute with $\v$, these
currents mix quark and anti-quark fields in contrast with the currents
associated to the flavor and spin symmetries:
$$
J ^{\mu} = \bar h_{v} \v v ^{\mu} h_{v} \;\;\; {\rm and } \;\;\;
J ^{\mu}_ v = \bar h_{v} v ^{\mu} h_{v}
\eqno(fc)$$
$$
J ^ {\mu} _ i= \bar h_{v} \v v ^{\mu} S _ i h_{v} \;\;\; {\rm and }
\;\;\; J ^{\mu} _ {vi} = \bar h_{v} v ^{\mu} S _ i h_{v} \, ,
\eqno(sc)$$
which contain either quark or anti-quark fields.
Since all the currents are multiplied by $ v ^\mu $, it is
convenient to define $j$ in such a way that $ j v ^\mu := J ^\mu $ for
all the currents
\(g5c)-\(sc). In the rest of the paper we will use $j$ when referring to
the currents.

In nature the flavor and spin symmetries are realized \`a la Wigner-Weyl
: $B$ and $B^{\ast}$ as well as $D$ and $D^{\ast}$ can be accommodated in
the dimension 4 representation of spin $SU(2)$ whereas $B$ and $D$ can
be accommodated in a dimension 2 representation of flavor $U(2)$. The
$U(1)$ factors just keep track of the separate quark and anti-quark
number conservation. No obvious $U(4)$ multiplet is observed which
suggests that the extra generators must be spontaneously broken in
nature. In the next section, we argue that the extra symmetries are
actually broken spontaneously.

\bigskip

{\bf 3. Realization of the $U(4)$ symmetry in the vacuum of the HQET}

 Firstly, we argue that the $\gamma _5 $-symmetry, given by \(g5),
is spontaneously broken using phenomenological information.

Consider a meson $M$ made up from a heavy quark $Q$ and a light
anti-quark $\bar q$. The meson decay constant $f _ M$ is defined by
$$
< 0 | \bar q \gamma ^{\mu} \gamma _5 Q | M _ {\bar q Q}
> = i f _ M p ^ {\mu}
\eqno(dc)$$
and is known from phenomenology to be different than zero.
Next consider the following matrix element
in the heavy quark approximation:
$$
< 0 | \bar q   \gamma ^ {\mu} h ^- _v  |a ^\dagger _ v  d ^\dagger
> \,  ,
\eqno(dch)$$
where by $a ^\dagger _ v \, , d ^\dagger $ we denote the creation
operators of
the heavy and light quarks correspondingly. Recall that $ h ^- _ v$
is the field which contains
the creation operator for the heavy anti-quark. $|0>$ denotes
the vacuum of the full QCD. The matrix element \(dch)
is zero, since the operator $ h ^- _ v $ annihilates the vacuum on the
left. Performing an infinitesimal $\gamma _5$-transformation
$$
\delta h ^ {\pm} _ v = i \gamma _5 \e \, h ^{\mp} _ v
\eqno(inf)$$
on the matrix element \(dch) we obtain
$$
<0 | \bar q \gamma ^{\mu} h ^- _  v | \delta a ^\dagger _ v \,
d ^\dagger > \, +\, \epsilon < 0 |
\bar q \gamma ^{\mu} i \gamma _5  h ^+ _ v | a ^\dagger _ v \,d ^\dagger
> \, +\, < \delta 0 | \bar q \gamma ^{\mu} h ^- _ v | a ^\dagger _ v
\, d ^\dagger > \, = \, 0
\eqno(del)$$
where $ \delta$ means variation with respect to $\gamma _5$.
The first term of the last expression is obviously zero, while the
second term is equal to the matrix element \(dc) in the HQET (up to
anomalous dimensions which are not important for the argument). Since
the meson
decay constant $f _ M $ goes like ${1 \over \sqrt m }$ for $m$ large
[2,7], $ f _ M p
^{\mu}$, that is the right hand side of \(dc), is different than zero
for $ m$ large. Then in order for the last
equation
to make sense the variation of the vacuum must be different from zero.
With this we conclude that the vacuum of the full QCD must not
be invariant under the $\gamma _5 $-symmetry.

The question now is whether the breaking of the $ \gamma _ 5$ generator
is due to non-perturbative QCD effects, as it is the case for the chiral
symmetry, or it can be understood in simpler terms. In order to
disentangle this issue let us analyze the transformation properties
of the vacuum of the HQET, when QCD is switched off, under the new
generators \(g5) -\(ev).

Let us begin by analysing (1) in first quantization. The equations of
motion for $h_{v}^{\pm}$ read
$$i\v v.\partial h_{v}=0 \quad \leftrightarrow \quad
i v \cdot \partial \, h ^{\pm} _ v = 0
\eqno(eq)$$
If we restrict ourselves to fields with well-defined energy, i.e.
$h_{v}^{\pm}(\vec x,t)=e^{ik_0 t}h_{v}^{\pm}(\vec x,0)$ we obtain the
following eigenvalue equation
$$
\hat H h_{v}^{\pm}=k_0 h_{v}^{\pm} \quad ;\quad
\hat H :=i{v^{i}\partial_{i}\over v^0} \eqno(hop)
$$ where $iv^{i}\partial_{i} / v^0 $ is the first quantized
Hamiltonian
operator, which is diagonal in the Dirac matrix space. The
eigenfunctions of $\hat H$ are plain waves of momentum $\vec k$ leading
to the dispersion relation
$$ k^0 = {\vec v \cdot \vec k \over v^0 } \eqno (dr) $$
Consider next the
infinitesimal $\gamma _5$-transformation \(inf).
Since the $\gamma _5$ matrix commutes with the Hamiltonian operator
$\hat H$, it follows that if $ h ^+ _ v $ is an eigenstate of $\hat H$
with energy $ k _ 0$ the transformed field $ i \gamma _5 h ^-_ v $
is also an eigenstate of $ \hat H $ with the same energy $ k _0$. The
same statement is true for any of the generators of the transformations
(6)-(9).

In the original Dirac theory the transformation \(inf) amounts to
taking a state of positive energy to a state of energy negative.
Indeed, the equations
$ ((1\mp \v ) /2) h ^{\pm} _ v = 0 $, which define the fields
$ h ^{\pm}_ v$ in this approximation, correspond to the positive and
negative energy solution respectively in the original theory.
In the massive Dirac theory, then, the present transformation
cannot be a symmetry, since in this theory
the positive energy states are separated from the negative energy states
by an amount of energy $ E \geq 2m $. Na\"\i vely, therefore, one would
expect
that in the infinite mass limit this mass gap becomes infinitely large
and no symmetry which would relate the positive with the negative energy
spectrum would exist. In the HQET, however, redefining the field
as $ \Psi (x) = e ^{-im \v v \cdot x } h _ v (x) $
we have removed the mass gap from the spectrum and
instead of infinitely separating the positive and negative sectors we
have brought them close together. Indeed, the extra time dependence of
the new
field is $ e ^{im v ^0 t } $ and the field redefinition amounts to
redefining the zero level of the energy spectrum by an amount equal to
$ m v ^0$. The correct picture of how the Dirac vacuum is modified in
this approximation is given in fig. (1).

It is worth noticing here that since the Hamiltonian operator \(hop)
is diagonal in the Dirac matrix space, we can multiply
a solution $ h _v ^{\pm}$ of the eigenvalue equation by any $4$x$4$
matrix and it still remains
a solution with the same energy and momentum. In other words, the
transformation of the $4$x$4$ matrices
is a symmetry of the equations of motion, and hence of the eigenvalue
equation \(hop) . If we associate solutions of the eigenvalue equation
\(hop) to one-particle states, then the $4$x$4$ matrix transformations
conserve the energy of the latter. If we further constrain these
transformations to preserve the
norm of the first quantized Hilbert space we obtain the $U(4)$ group.
This is the physical argument promised before. The
invariance of the action (1) restricts automatically these $4$x$4$
matrices to those of the non-unitary representation of $U(4)$, as
explained in the previous section.

The Hamiltonian of the system can be read off (1)
$$
H = -\int d^3\vec x (i \bar h ^+ _ v v ^i \partial _ i h ^+ _v \; - \; i
\bar h ^- _ v v ^i \partial _ i h ^-_ v )
\eqno(1h)$$
where $\pm i\bar h_{v}^{\pm}v^0$ are the canonical momenta of
$h_{v}^{\pm}$.
At the second quantized level the field variables $h ^{\pm} _ v$ are
most conveniently expressed in terms of annihilation and creation
operators. They are given by
$$
h ^+ _ v (x) = \int {d ^3 \vec k \over (2 \pi ) ^3 v^0} \sum _
{\sigma = 1,2}
u ^\sigma _ v  a ^\sigma _ v ({\vec k})
e ^{-ik \cdot x} \;
$$
$$
h ^- _ v (x) = \int {d ^3 \vec k \over (2 \pi ) ^3 v^0} \sum _
{\sigma = 1,2}  v ^\sigma
_ v  { b ^\dagger} ^\sigma  _ v ({\vec k}) e ^{ik \cdot
x} \; ,
\eqno(ho)$$
where $ a^\sigma _v ({\vec k}) $ annihilates heavy quark
and $ {b ^\dagger} ^\sigma _v ({\vec k})$ creates heavy anti-quark
respectively of small momentum ${\vec k}$ about $m v ^\mu  $. The
constant Dirac spinors $u ^\sigma _ v $ and $ v ^\sigma _ v  $
are taken with the following normalization:
$$
\eqalign{ {\bar u} ^ \sigma _ v  u ^ {\sigma ^ \prime } _ v
 &=  \delta _ { \sigma \sigma ^ \prime} \, ,
\;\;\; {\bar v} ^ \sigma _ v  v ^{\sigma ^ \prime } _ v
= - \delta _ {\sigma \sigma ^ \prime} \cr
{\bar u} ^ \sigma _ v  v ^{\sigma ^ \prime } _ v
 &= 0 \, ,
\;\;\;\; {\bar v} ^\sigma _ v  u ^{\sigma ^\prime } _v
 = 0  . \cr }
\eqno(nor)$$

Using the anticommutation relations of the fields $h ^{\pm} _v $,
given by
$$
\{ h ^{\pm} _ v (\vec x) , \bar h ^{\pm} _ v (\vec y) \} \,= \,
{\pm} ({1 {\pm} \v \over 2}){1\over v^0}
\, \delta ^3 (\vec x-\vec y) \;\;\;
\eqno(crh)$$
we obtain the following anticommutation relations for the creation and
annihilation operators:
$$
\{ a ^\sigma _v ({\vec k}) , { a ^\dagger }^{\sigma ^\prime}_ v
({\vec k} ^\prime )\}\, = \, v ^0 (2\pi)^3 \delta ^3 ( {\vec k} - {\vec
k} ^\prime) \, \delta _ {\sigma \sigma ^\prime}
\,\,\, {\rm and} \,\,\, \{b^\sigma _v ({\vec k}) , {b ^\dagger}
^{\sigma ^\prime}_v ({\vec k} ^\prime ) \} = \, v ^0 (2\pi)^3
\delta ^3 ({\vec k} - {\vec k} ^\prime)
\, \delta _ {\sigma \sigma ^\prime}
\eqno(cr)$$
All other anticommutation relations are zero.

Substituting now the expressions \(ho) into \(1h) we
obtain for the second quantized Hamiltonian$$
H = - \int {d ^3 \vec k \over (2 \pi ) ^3 v^0} \sum _{ \sigma =1,2} { v
^i k _ i \over v ^0 } \; [ \;{a ^\dagger } ^\sigma _v ({\vec k}) \,
a ^\sigma _ v({\vec k}) \,+\, {b ^\dagger} ^\sigma _ v ({\vec k} )\, b
^\sigma _ v ({\vec k}) \;] \; .
\eqno(ham)$$
The first thing to notice about this Hamiltonian is that its spectrum
is unbounded from both above and below.
Indeed, since the momentum fluctuation $k ^i $ can take
any value around zero the Hamiltonian \(ham) can be negative.
However, the full Hamiltonian is always positive definite.
The full energy of the system $ E = \sqrt { m ^2 + {\vec k}^2 }$ takes
the form $ m v ^0 + { \vec v \cdot \vec k \over v ^0 } $ in first
order in the
$1/m$ expansion, where  $ v ^i k _ i $ is much smaller than the $ m v
^0$. In the HQET we count the energy of the states above the value
$m v ^0$ and the effective
Hamiltonian \(ham) is a small correction of the full Dirac
Hamiltonian. Its negative value does not disturb the positiveness
requirement of the latter.

Next we build the Hilbert space starting from the vacuum of the theory
defined as the state which is annihilated from both $a^\sigma _ v ({\vec
k})$ and $b ^\sigma _ v ({\vec k})$. It is given by
$$
a ^\sigma _v ({\vec k}) \, b ^\sigma _v ({\vec k})\; | 0 ; 0> = 0 \, .
\eqno(vd)$$
The Hilbert space is a direct product of the Hilbert space of quarks
and the Hilbert space of anti-quarks and we denote the vacuum by $
|0;0>$. Excited states are denoted by the momentum and helicities of
their quarks and anti-quarks. This is a convenient notation since in
this theory quarks and anti-quarks coexist and each can occupy states of
either positive or negative energy (see fig. 1 and 2).

The so defined vacuum has zero energy
$$
 H \;|0; 0> = 0.
\eqno(ve)$$
Notice, however, that this is not a state of minimum
energy in the effective theory. This is only the  `vacuum' in the sense
that it corresponds to the vacuum of the original Dirac theory.
All excited states can now be constructed, as usually, by operating with
the creation operators on the vacuum.

The transformation $\(g51)$ in the second quantized picture
takes the following form
$$
\eqalign{ a ^\sigma _v ({\vec k}) &\rightarrow cos \e \;
 a ^\sigma _v ({\vec k}) \,+\, \sum _ {\sigma ^\prime } i \bar
u ^{\sigma } _v  \gamma _5 v ^{\sigma^{\prime}}  _ v
\,sin \e \; { b ^\dagger } ^{\sigma^{\prime}}_ v (-{\vec k}) \cr
b ^\sigma _v ({\vec k}) &\rightarrow cos \e \; b ^\sigma _v  ({\vec k})
+ \sum _ {\sigma ^\prime} i \bar u ^{\sigma^{\prime}} _v   \gamma _5 v
^{\sigma }
_v  sin \e \;{ a ^\dagger } ^{\sigma^\prime}_v (-{\vec k}) \, . \cr }
\eqno(tr2)$$
The expression for the creation operators are obtained by taking the
hermitian conjugate in equation \(tr2).

The charge operator associated to the symmetry transformation (6) reads
(from (17))
 $$
Q = \int d ^3 \vec x  ( \,\bar h ^+ _ v  v ^0 \gamma _5 h ^- _ v  \,-\,
\bar h ^- _ v v ^0 \gamma _5 h ^+ _ v \,)\; ,
\eqno(ch)$$
which expressed in terms of creation and annihilation operators takes
the form
 $$
Q = \int {d ^3 \vec k \over (2 \pi) ^3 v^0} \sum _ {\sigma ,  \sigma
^\prime} \;[ \,\alpha ^{ \sigma   \sigma ^\prime} _ v  \,
{a ^\dagger}^\sigma _v ({\vec
k}) {b ^\dagger} ^{\sigma ^\prime} _ v (-{\vec k}) \,-\, \beta ^{\sigma
 \sigma ^\prime}
_v  \, b ^\sigma _v (-{\vec k}) a ^{\sigma
^\prime} _v ({\vec k}) ] \; ,
\eqno(ch2)$$
where $ \alpha ^{\sigma \sigma ^\prime } _v  = \bar
u ^\sigma _ v   \gamma _5 v ^{\sigma ^\prime}
_v  $ and $ \beta ^{ \sigma \sigma ^\prime} _ v
= \bar v ^{\sigma  ^\prime} _v  \gamma _5 u ^\sigma _ v
 $. It is, then, straightforward to show that
$$
[ Q ,H ] = 0
\eqno(cq)$$
where we have used the anticommutation relations \(cr).
Notice next that by acting with expression \(ch2) on the vacuum we
obtain $$
Q |0;0> = \int { d ^3 \vec k \over (2 \pi ) ^3 v^0} \sum _ {\sigma ,
\sigma ^\prime } \alpha ^{\sigma \sigma ^\prime} _v  \;
|{\vec k} , \sigma ; -{\vec k} , \sigma ^\prime >\, \ne \, 0 \, .
\eqno(1s)$$
This shows that the charge operator does not annihilate the vacuum.
The meaning of this is that
the symmetry is not realized \`a la Wigner-Weyl but \`a la
Nambu-Goldstone, i.e. there is spontaneous
symmetry breaking. We devote the rest of the section to
analyzing this phenomenon.

The symmetry (6) is realized on the Hilbert space by the
unitary operator
$ U = e ^{i \e Q} $ (recall that $Q$ is hermitian). The fact that $U$
is unitary guarantees that the symmetry transformations preserve the
normalization of the states even if this symmetry is spontaneously
broken. We then have
$$
\eqalign{&U |0;0>= |0 ;0>  + i  \e \int {d ^3 \vec k \over (2 \pi ) ^3
v^0} \sum _ {\sigma , \sigma ^\prime } \alpha ^{\sigma \sigma ^\prime}
_v
 |{\vec k} , \sigma ; -{\vec k} , \sigma ^\prime > \cr
&+ {\e ^2\over 2} \int d ^3 \vec k \delta^3( 0)
\sum_{\sigma , \sigma^{\prime}} \beta
^{\sigma \sigma ^\prime} _ v  \alpha ^{\sigma^{\prime}
\sigma  } _ v  |0;0>  \cr
&- {{\e} ^2\over 2}
\int \int { d ^3 \vec k \over (2 \pi ) ^3 v^0} {d ^3 \vec k _1 \over (2
\pi)
^3 v^0} \sum _ {\sigma, \sigma ^\prime , \sigma _ 1 , \sigma ^\prime _
1} \alpha ^{\sigma  \sigma  ^\prime }_ v \alpha
^{\sigma  _1  \sigma ^\prime _ 1}  _ v
 |{\vec k}, \sigma, {\vec {k _ 1}},
\sigma _1 ; -{\vec
k} , \sigma ^\prime ,-{\vec{ k_1}} ,\sigma ^\prime _ 1> +...  \cr }
\eqno(uo)$$
The meaning of the last expression is that
the $\gamma _5 $-symmetry operator acting on the
vacuum gives states with so many quarks of momentum $\vec k$ as many
anti-quarks of momentum $-\vec k$. All these states have
energy zero since a quark of momentum $ \vec k$ adds energy equal
to $ v^ i k _i /v^0$ in the system and an anti-quark of momentum
$-\vec k$ adds an amount of energy equal to $- v ^i k _i/ v^0$.
This shows that the vacuum of the HQET is infinitely
degenerate and the $\gamma _5$-symmetry mixes these degenerate states
between them. A picture of how all these degenerate vacua look is
given in fig. 2.

This degeneracy of the vacuum is, then, the mechanism through which
the breaking of the symmetry manifests itself in this formalism. The
symmetry is spontaneously broken whenever a specific vacuum state is
chosen. This leaves us with the question of how this result reconciles
with the Goldstone theorem. The answer to this question lies in the
non-relativistic nature of the theory and it is analyzed in the next
section.

Notice at this point that we have provided a simple theoretical
explanation in terms of the free theory (no gluon fields) of
the phenomenological observation made
at the beginning of this section by which the $\gamma_5$-symmetry must
be spontaneously broken in nature.

The analysis above extends trivially to the rest of the
transformations (7)-(9). That is, all these symmetries are spontaneously
broken. On the other hand the normal ordered charges
corresponding to
the flavor and spin symmetries (2) and (3) do annihilate the vacuum and
hence they are relized \`a la Wigner-Weyl.

\bigskip

\bigskip {\bf 4. Goldstone modes}

Before going into the physical implications of having a symmetry
spontaneously broken in the HQET, let us briefly recall what is
generically known for such a situation.
Whenever we have spontaneous symmetry breaking in a local quantum field
theory with short range interactions, the Goldstone theorem applies. For
relativistic theories, it implies that massless particles arise in the
spectrum [15]. For non-relativistic theories the theorem is less
restrictive. It only implies that there exist collective excitations
such that their energy vanishes when their momentum does so exist.
The particular way in which the energy vanishes, i.e., the dispersion
relation at low momentum, depends however on every particular theory.
It is not fixed to be $E=\mid\vec k\mid $ by Lorentz covariance as it is in
relativistic theories. (See [16] for a discussion in condensed matter
physics and [17] for a general discussion.)

Since the HQET is a non-relativistic theory, we should not expect any
detailed information on the low momentum dispersion relation from the
Goldstone theorem. A Goldstone mode of momentum $p$ is created when
the Fourier transform of the time component of the current associated to
the broken generator acts on the vacuum.

In order to study the Goldstone modes in this theory, we firstly analyze
the properties of the currents \(g5c) and \(ec) corresponding to the
broken symmetries under the unbroken generators. We
have a freedom in
the choice of the currents since any linear combination of them is also
conserved. We choose the
following combination:
$$
  j _{5\pm}:=\bar h_{v}i \gamma_5 p_{\pm} h_{v}  \quad
{\rm and}\quad
 j _{5\pm}^{i}:=\bar h_{v}i {\el } _{i} p_{\pm} h_{v} \, ,
\eqno(bc)$$
where by $ p _ {\pm} $ we denote the projection operator $ (1{\pm} \v
)/ 2 $. This suitable combination
of currents corresponding to the spontaneously broken generators
can be accommodated into two dimension 4 irreducible representations
of $U(2)\otimes U(2)$. Indeed, they transform as follows under the
unbroken flavor and spin symmetries:
 $$\eqalign{
\delta_{\theta_{\pm}}j_{5+}=\pm \theta_{\pm}i j_{5+} & \quad\quad
\delta_{\theta_{\pm}} j_{5-}=\mp \theta_{\pm}i j_{5-} \cr
\delta_{\e _{\pm}^{i}}j_{5+}=
\mp\e_{\pm}^{i} 4 i j_{5+}^{i} &\quad\quad
\delta_{\e _{\pm}^{i}}j_{5-}=
\mp\e_{\pm}^{i}  4 i j_{5-}^{i} \cr
\delta_{\theta _{\pm}}j_{5+}^{j}=\pm \theta_{\pm} i
j_{5+}^{j} &\quad\quad
\delta_{\theta_{\pm}}j_{5-}^{j}=\mp \theta_{\pm} i
j_{5-}^{j} \cr
\delta_{\e_{\pm}^{i}}j_{5+}^{j}=
4\e_{\pm}^{i}\epsilon^{ijk}j ^k _{5+} \mp \e_{\pm}^{i} i
j_{5+} &\quad\quad
\delta_{\e_{\pm}^{i}}j_{5-}^{j}=
4\e_{\pm}^{i}\epsilon^{ijk}j_{5-}^{k} \mp \e _{\pm}^{i} i
j_{5-}  }
\eqno (50)
$$
where the flavor and spin parameters of the transformations $
\theta _ {\pm}$ and $ \e ^i _ {\pm}$ are defined as
in \(fl) and \(sp).
The linearity of the last transformation asures that this is a proper
representation of the full unbroken subgroup acting on the 8-dimensional
space of the currents corresponding to the broken generators. Since,
however, the $+$ and
$-$ sectors of the broken generators factorize under the action of the
unbroken ones, this representation is reducible. The 2 irreducible
representations are now 4-dimensional. The Goldstone space is spanned
by the broken currents.

Next we construct the Goldstone modes corresponding to the currents
\(bc). For this purpose and what follows it is convenient to
introduce the following notation:
$$
j _ {\Gamma ^A _{\pm}}:= \bar h_{v} \Gamma ^A _{\pm} h_{v}\,,
\quad\quad {\rm where} \quad
\Gamma ^A _{\pm}=i\gamma_5 p_{\pm} , \; i{\el}_i p_{\pm} \, .
\eqno(not)$$
We generically consider the Fourier transform of the time component
of the current $J _ {\Gamma ^A  _ -}$ acting on the vacuum.
It is given by
$$
\int d ^3 {\vec x} e ^{i {\vec p } \cdot {\vec x}} J ^0 _ {\Gamma ^A _
-} (x) | 0;0> \,=\,
\int {d^3 \vec k \over (2 \pi) ^3 v^0} \sum_{\sigma ,
\sigma ^{\prime}}\alpha_{v}^{\sigma\sigma^{\prime}}
 \vert \vec k , \sigma ; -\vec k- \vec p , \sigma^{\prime}> \, ,
\eqno(ftg)$$
where we have substituted \(ho) in the last step of \(ftg).
Here $ \alpha _ v ^{\sigma {\sigma ^\prime}}  = \bar u _ v
^\sigma  v ^0 \Gamma ^A _ - v _ v ^{\sigma ^\prime} $.
It is now straightforward to indentify the last expression as
a state whose energy goes to zero (being the energy
of a heavy quark of momentum ${\vec k} $ and of a heavy anti-quark  of
momentum $- {\vec k }$) whenever the spatial momentum $ {\vec p}$ goes
to zero. This is consistent with the Goldstone theorem for
non-relativistic theories.

Notice at this point that only the $J ^0 _ {\Gamma ^A_ -}$ components
of the broken currents can create a Goldstone state when acting on the
vacuum ($J ^0 _ {\Gamma ^A _ +} $ annihilate the vacuum). Hence, only
four independent Goldstone modes can be created in this theory. This
result is related to the fact that the irreducible representations are
4-dimensional.

Let us next take a simple minded point of view and suppose that this
picture holds
even when the full QCD is switched on. This is very plausible at the
level of the HQET since the gluons are blind to all symmetries
\(fl1)-\(sp1) and \(g5)-\(ev). Since the Goldstone states
contain a heavy quark and a heavy anti-quark, they may be identified
with $\bar c c$ or $\bar b b$ states where
the $2m v ^0$ energy dependence corresponding to the mass of the heavy
quark and anti-quark has been removed. They would correspond to the
$\eta_{c}$ , $J/\Psi$ or $\eta_{b}$ , $\Upsilon$ particles
for the $c$ and $b$ quark respectively. $J ^0 _{5-}$ would create the
pseudoscalar mesons whereas $J ^{0i} _{5-}$ the vector mesons.
This, then, would have the immediate physical consequence that
 $\eta_{c}$ and $J/\Psi$ must have the same mass, since under the
symmetry transformations they belong to the same multiplet. (The same
holds
for $\eta_{b}$ and $\Upsilon$.) Current data tells us that the former
fit the bill quite well. ($\eta_b$ has not been found yet.)

Let us emphasize at this point that the identification of
$\bar c c$ or $\bar b b$
states with the Goldstone modes must be done {\it after} the $2mv^0$
dependence of the energy on the heavy quark masses is substracted. In a
non-relativistic theory a Goldstone mode does not mean a zero mass
particle. In our case, it only means that the residual energy of the
$\bar c c$ or $\bar b b$
states, that is the energy once the the $2mv^0$ has been substracted,
must go to zero when the three momentum goes to zero (see the discussion
below). Whether this actually occurs or not in nature is a separate
question we shall comment upon in the last section.

In order to find the dispersion relation of the
Goldstone mode, it is enough to calculate the following current-current
correlator in the HQET:
$$
\Pi_{\Gamma ^A _\pm}^{HQET}(p):=\int d^4 x  e^{ip \cdot x} <0 ; 0\vert
T(j _{\Gamma^A _ \pm}^{\dagger}(x) j_{\Gamma ^A _ \pm}(0))\vert 0; 0>
\eqno (47) $$
We obtain
$$ \Pi_{\Gamma ^A _ \pm}^{HQET}(p)=-\int {d^4 k\over (2\pi)^4}
tr(\bar\Gamma ^A _{\pm}\Gamma ^A _{\pm})
{i \over \mp v \cdot k+i\epsilon }\; {i \over \pm v \cdot (k-p) +
i\epsilon} \quad\quad
\eqno (48) $$
where by $ \bar \Gamma ^A _ {\pm} $ we denote $ \gamma ^ 0 ({\Gamma ^A
_ {\pm} }) ^\dagger \gamma ^ 0 $. The last expression
is ill-defined. However, using a spherical cut-off in
the $\tilde k_{i}= k \cdot e_{i}$ space, which respects all the
symmetries of the HQET, at least when QCD is switched off, it becomes
$$ \Pi_{\Gamma ^A _ \pm}^{HQET}(p) =-i
tr(\bar\Gamma ^A _{\pm}\Gamma ^A _{\pm} )
{1\over \mp v \cdot p+i\epsilon} \int^{\Lambda}{d^3 \tilde {k^{i}}
\over (2\pi)^3}
\eqno (49) $$
which is well-defined. Notice that $
tr(\bar\Gamma ^A _{\pm}\Gamma ^A _{\pm} ) $
is negative definite and
hence the residu of the pole has the right sign. The cut-off dependence
can be removed by a wave function renormalization
of the current.\footnote{$^{\dagger}$}{If the integrals \(48) and \(49)
are evaluated by
using dimensional regularization, the result is zero (since there is no
scale in \(49)). Dimensional
regularisation however breaks the $U(4)$ symmetry and hence it is not
suitable to address questions which are intimately related to this
symmetry.}

The meaning of the last expression is that the correlator \(47) has a
pole at $ v \cdot p = 0 $, which corresponds to the possibility of
virtual exchange of this Goldstone mode.
The dispersion relation of this mode is identical to the one of the
fields $h_{v}^{\pm}$ appearing in the lagrangian of the HQET (1).
The dispersion relation that we find for the Goldstone mode is then the
generic dispersion relation for a field describing heavy particles
\(dr), the leading mass dependence of which has been removed [7,18].
Notice that since the energy of this mode goes to zero when the
momentum does so (i.e., $p ^0 = {{\vec p} \cdot {\vec v} \over v ^0 } $) it
is perfectly compatible with the Goldstone theorem for non-relativistic
theories. If this simple minded picture is correct one should be able
to apply to this case all the
well-known machinery of low energy effective lagrangians for Goldstone
bosons [19] (see also [20]).

Before closing this section we would like to point out the role of the
$i \e$ in the propagators \(48) as a symmetry breaking parameter. If the
$i \epsilon$ prescription is incorporated in the lagrangian, this
becomes
$$
L = \bar h _ v ( i \v v \cdot D \,+\, i \e ) h _ v \, .
\eqno(le)$$
Therefore, it amounts to an infinitesimal source which breaks the U(4)
symmetry. The result of the current-current correlator,
which is formally covariant under the U(4) transformations
without the inclusion of the $i\e$, becomes well
defined only when the $ i \e$ is included and the limit $ \e
\rightarrow 0$ is taken. This is a well-known fact of the symmetry
breaking phenomenon, which manifests itself in this subtle way in our
calculation of the current-current correlator.
\bigskip
{\bf 5. Connection with the fundamental theory}

In order to have a more precise interpretation of the physical meaning
of the currents corresponding to the broken generators, it is convenient
to have their representation at the level of the fundamental theory. By
na\"\i vely applying the HQET rule, that is, redefining the heavy quark
field as
$$ \Psi \longrightarrow e^{-i\v mv \cdot x} h_{v} $$
we obtain the following expressions
$$ j _{\Gamma ^A _ {\pm}}=  \bar \Psi \Gamma ^A
_{\pm} \Psi e^{{\pm} 2i mv \cdot x}
\eqno (51) $$
for the currents \(not) in terms of the original field $\Psi$.
Below we present a more
careful derivation of this connection which in fact brings in some new
features. Notice that \(51) are local sources with suitable momentum
insertions.

Let us next consider the Dirac lagrangian with the most general
bilinear sources, given by
$$
L =  \bar \Psi(i \Dsl -m )\Psi + S \bar \Psi \Psi + P \bar \Psi
\gamma
_ 5 \Psi + V _{\mu} \bar \Psi \gamma ^ \mu \Psi + A _{\mu} \bar \Psi
\gamma ^ \mu \gamma _ 5 \Psi + T _{\mu \nu} \bar \Psi \sigma ^ {\mu \nu}
\Psi
\eqno(dd)$$
Then since our symmetry generators $i p ,
i \gamma _ 5 p _ {\pm} , i \el _ i p _ {\pm} , i S _ i p _ {\pm} $
 form a basis of the 4x4 matrices (see discussion on section 2),
the last lagrangian can be written as
$$
\eqalign{L=  \bar \Psi(i \Dsl -m ) \Psi  &
+ i \bar \Psi \gamma_5 p _ +
\Psi A  _{5 +} + i \bar \Psi \el_ j p _ + \Psi A ^j
_ {5 +} + i \bar \Psi S _ j p _ + \Psi A ^ j _
+ +i \bar \Psi p _ + \Psi A _ + \cr
&+ i \bar \Psi \gamma_5 p _ -
\Psi A  _{5 -} + i \bar \Psi \el_ j p _ - \Psi A ^j
_ {5 -} + i \bar \Psi S _ j p _ - \Psi A ^ j _
- +i \bar \Psi p _ - \Psi A _ - \cr }
\eqno(Dfl)$$
where we have defined
$$
\eqalign {A _ {\pm} &= -i S \, \;{\mp}\, \;i V _ {\mu} v ^{\mu} \cr
A _ {5 \pm} &= -i P \,\;{\pm} \,\;i A _ {\mu} v ^{\mu} \cr
A ^i _ {\pm} &= - {1 \over 2} T _ {\mu \nu } \e ^{ijk} e ^{\mu } _ j e
^{\nu } _ k \,\;{\pm} \,\;{i \over 4 }A _ {\mu} e ^{\mu} _ i \cr
A ^i _ {5{\pm}}&= i V _ {\mu} e ^{\mu} _ i \;\, {\pm}\,\; 2i T _ {\mu
\nu } e ^{\mu} _ i v ^{\nu} \cr } \,.
\eqno(aa)$$

Next if we restrict the last sources such that they are given in terms
of the slowly varying $a_{5{\pm}}$ and $ a ^j _{5{\pm}}$, as
$$
\eqalign{A_{5{\pm}}&=a_{5{\pm}} e^{{\pm} i2mv \cdot x} , \;\;\; A ^j_ {
{\pm} }= a ^j _ {{\pm}} \cr
A ^j _{5{\pm}}&=a^j _{5{\pm}} e^{{\pm} i2mv \cdot x} , \;\;\; A _  {\pm}
= a _  {\pm} \cr }
\eqno(sor)$$
because of \(51), the current-current correlators of the HQET
such as the \(47), can be generated from the fundamental theory
defined by \(Dfl). Indeed, following a derivation similar to ref. [13]
we obtain
$$
Z( a ^A , a ^V ,\bar\eta_{v},\eta_{v})={det \DI \over det {\DI}_{v}}
 Z_{\rm HQET}(a ^A ,a ^V,\bar\eta_{v},\eta_{v})
\eqno(ex)$$
where
$$\eqalign{Z&= \int D \bar \Psi D \Psi e ^{i \int d ^4 x ( \bar \Psi \DI
\Psi + \bar \Psi \eta + \bar \eta \Psi )  }\cr
Z_{HQET}&= \int D \bar h _ v D h _ v e ^{i \int ( \bar h _ v {\DI} _ v h
_ v + \bar h _ v \eta _ v + \bar \eta _ v h _ v ) } \cr
\DI &=  i \Dsl -m  + a ^V _ + \Gamma ^V _ + + a ^V _ - \Gamma ^V _ - +
a ^A _ + \Gamma ^A _ + e ^{2i m v \cdot x }  + a ^A _ - \Gamma ^A _ -  e
^{-2im v \cdot x}  \cr
{\DI} _{v}&= i \v v \cdot D + a ^V _ + \Gamma ^V _ + + a ^V_ - \Gamma ^V
_ - + a ^A _ + \Gamma ^A _ + + a ^A _ - \Gamma ^A _ -  \cr}
\eqno(def)$$
and we have defined
$$
 a ^A _ {\pm} = (a _ {5 {\pm}} , a ^j _ {5 {\pm}}) ,\;\;\;
 a ^V _ {\pm} = ( a _ {\pm}, a ^j _ {\pm} ), \;\; {\rm and } \;\;
 \eta _ v =  e ^{i m \v v \cdot x } \eta   , \;\;\;
 \bar \eta _ v = \bar \eta e ^{ -i m \v v \cdot x }\, .
 $$
$ \Gamma ^A _ {\pm} $ is given as in \(not) and $ \Gamma ^V _
{\pm} = i p _ {\pm} ,  i S _ i p _ {\pm}$.
In deriving the expression \(ex) the $m \rightarrow
\infty$ limit has been taken in the $Z _ {\rm HQFT} $ and the
$det \DI _ v$, while the mass dependence of $det \DI$ is kept. The
latter is meant to be calculated in the large $m$ limit as well.
Integration over gluons and light quarks must be understood in \(ex).
The only approximation made in obtaining this expression is that slowly
varying (soft) gluons
dominate completely the path integral so that one can always consider
$B_{\mu\nu}/m^2<<1$. In reality, hard gluons are certainly not
negligible. However, since QCD is an assymptotically free theory their
effect can be accounted for in perturbation theory and it only amounts
to the incorporation of suitable anomalous dimensions to the currents
[2,6,7]. We shall disregard it in this work.

 From \(ex) we see that $Z=Z_{\rm HQET}$ except for the ratio of
determinants.
Here this ratio is non-trivial. On one hand, because of the
oscillating exponentials in $\DI$ , $det \DI$ does not admit
a local derivative expansion for large $m$ (it is
non-analytical at low momentum).
On the other hand, because of the terms of the type $\Gamma^{A} a^{A}$ ,
$det{\DI}_{v}$ is not a constant and does not admit a local derivative
expansion either. In fact they give qualitatively different
contributions so there is no {\it a priori} reason why they should be
neglected. This can be seen, for instance, by calculating
the correlator corresponding to \(47) in terms of the original fields
$\Psi$ when QCD is switched off. It is given by
$$  \Pi_{\Gamma^{A}_{\pm}}(p)=\int d^4 x e^{ip \cdot x} <0\vert T(\bar
\Psi
(x) \bar\Gamma ^A _{\pm} \Psi (x) \bar\Psi(0) \Gamma ^A _{\pm} \Psi(0))
e^{ {\mp} i2mv \cdot x} | 0> \, .
 \eqno (52) $$

This expression, except for the exponential factor,
describes a one loop
fermion diagram and for finite fermion mass has a branch point singularity
corresponding to a pair creation and not a pole singularity as found
in \(49). The branch point singularity at the value of the external
momentum $ q ^2 = 4 m ^2$ amounts to a non zero contribution in the
imaginary part of the matrix element starting from this value of the
momentum. In the large m limit with the external momentum $q ^\mu$
fixed we expect zero contribution from this diagram, since the
creation of two heavy particles requires infinite external energy.

The result obtained from \(52) for large $m$, when we have swiched off
QCD, is
$$
\eqalign{Im(i\Pi_{\Gamma^{A}_{\pm}}(p)) &= {1\over 2}
tr(\bar\Gamma ^A _{\pm}\Gamma ^A _{\pm} ) \left(
 m^2 {1\over 2
\pi}\sqrt{\mp {v \cdot p\over m}}
 \mp m {1 \over 16 \pi} {p ^2 \over {v \cdot p } } \sqrt{\mp { v \cdot p
\over m}} \right) \cr
&+
%{\rm \; local\; counterterms} +
{\rm subleading \; terms\; in}\; m \cr }
\eqno (53) $$
which does not have a well-defined limit when $m \rightarrow \infty$.

The last result may seem paradoxical but it can be understood by
looking at the expression \(52). The exponential factor in this
expression amounts to putting heavy particles in the external lines
(equivalently the external momentum $q ^\mu $ takes the value
$  p ^\mu \mp 2m v ^\mu $, where $p ^\mu$ is very small).
Thus, in the large $m$ limit the external momentum is enough to produce
a pair and in principle one would expect to have a branch point
singularity in the $ \Pi _{\Gamma^{A}_{\pm}}(p ) $ given by \(52).
By taking, however, $p ^\mu$
small we are making an expansion exactly on the singular point. Although
for $m$
large the singularity moves to infinity, our
expansion moves with it and thus it retains a big mass dependence.
Notice that the expression \(53) corresponds to an imaginary part
only for $\mp v \cdot p /m > 0$.
This simply reflects the fact that for values of $q^2$ below the
threshold there is no contribution from the imaginary part of the
correlator (57) to the matrix element.
Indeed, when we take $q^{\mu}=p^{\mu}\mp
2mv^{\mu}$ for $p^{\mu}$ small we obtain $q^2 \sim 4m^2 (1\mp v\cdot
p /m) $.

We conclude then that the correlation functions of operators of the type
\(51) do not have a smooth large $m$ limit to a local HQET describing
particles and antiparticles with the same velocity. In spite of this,
let us point out some interesting features of the determinants in \(ex)
(keeping QCD switched off). At second order in the currents we have
$$ \eqalign{ tr log \DI &=  tr log \DI ^0
- {1\over 2} tr [ \,({\DI ^0}) ^{-1} \,
(\Gamma ^V _ {+} a ^V _ {+} +
\Gamma ^V _ {-} a ^V _ {-})
\,({\DI ^0})^{-1} \,
(\Gamma ^V _ {+} a ^V _ {+} +
\Gamma ^V _ {-} a ^V _ {-})
\,]\cr
&- {1\over 2} tr [
\, {(\DI ^0)}^{-1} \,\Gamma ^A _ {+} a ^A _ {+} e
^ {2im v \cdot x}\, ({\DI ^0}) ^{-1} \,\Gamma ^A _ {-} a ^A _ {-}
e^ {-2im v \cdot x} \, \cr &
+
\, {(\DI ^0)}^{-1} \,\Gamma ^A _ {-} a ^A _ {-} e
^ {-2im v \cdot x}\, ({\DI ^0}) ^{-1} \,\Gamma ^A _ {+} a ^A _ {+}
e^ {2im v \cdot x} \,
] \cr}
$$
$$
\eqalign{tr log \DI_{v} &=  tr log \DI ^0 _ v
- {1\over 2}tr [\, ({\DI _ v ^0}) ^{-1} \,
(\Gamma ^V _ {+} a ^V _ {+} +
\Gamma ^V _ {-} a ^V _ {-})
({\DI _v ^ 0}) ^{-1} \,
(\Gamma ^V _ {+} a ^V _ {+} +
\Gamma ^V _ {-} a ^V _ {-})
\,] \cr
& - {1\over 2} tr[\,
 ({\DI_v ^0}) ^{-1} \,
\Gamma ^A _ {+} a ^A _ {+}
\, ({\DI _ v^ 0}) ^{-1} \, \Gamma ^A _ {-} a ^A _ {-} \, \cr &+
 ({\DI_v ^0}) ^{-1} \,
\Gamma ^A _ {-} a ^A _ {-}
\, ({\DI _ v^ 0}) ^{-1} \, \Gamma ^A _ {+} a ^A _ {+} \,
] \cr }
\eqno (trtr) $$
where $\DI ^0 = i \Dsl - m $ and $ \DI _ v ^0 = i \v v \cdot D $.
The linear terms in the sources which would appear in the expansion of
$tr log \DI $  are dropped because they contain oscillating
exponentials that cannot cancel between themselves and hence they do not
contribute [13].
Exactly the same occurs with quadratic terms containing $a ^A _ {\pm}$
twice.

 For the case of the $ det \DI_{v}$ the terms involving $\Gamma^{V}$
type of currents give zero after the integration over the momentum,
since the two poles of the two propagators appear in the same complex
half plane. The analogous terms for the $ det \DI$ reduce to
local counterterms and we shall disregard them. Let us then
concentrate on the terms containing $\Gamma^{A}$ type of currents.
{}From the previous results \(49) and \(53) we see that these terms
have different singularity structure. However, they do
have the same structure as far as the $U(4)$ symmetry is concerned.
Indeed, except from the exponential factors in $ \DI $, the terms which
contribute in both determinants are the same combinations of $+$ or $-$
components in the currents. Furthermore they are both invariant
under the unbroken subgroup $U(2)\otimes U(2)$. Therefore one could use
the fact that, even if $det \DI /det {\DI}_{v}$ is different from one
(and non-local), symmetrywise the properties of $Z$ are the same as the
properties of $Z_{\rm HQET}$. The key question is whether this nice
feature survives when QCD is switched on. Unfortunately, it is not
difficult
to convince oneself that in this case the expansion $B_{\mu\nu}/m^2<<1$,
i.e., the usual expansion in this framework, is singular. (The term
proportional to $ B _ {\mu \nu} B ^{\mu \nu} $ has a singularity of
higher order than the one in \(53) and terms of higher orders in the
gauge field have even higher order singularities.)
This means that
infrared QCD effects are crucial to answer this question and hence it
cannot be addressed in a reliable way by using the standard derivative
expansion
techniques [8,13] . Speculation on this issue is left to the following
section.

\bigskip
{\bf 6. Summary and discussion}

We have pointed out that the HQET describing a quark and an anti-quark
with the same velocity, as first written down in [5], enjoys an
invariance larger than the known flavor and spin symmetry. The
extra symmetries are of axial type and mix quarks and anti-quarks.
The full invariance of the theory for a single flavor corresponds to a
$U(4)$ group.

We have, then, given a phenomenological argument
to show that this symmetry must be spontaneously broken in nature.
We have also shown at the level of the HQET that the symmetry breaking
takes place even when QCD is neglected. Consequently, unlike chiral
symmetry breaking, it must not be regarded as a non-trivial feature of
the QCD vacuum but rather as an intrinsic feature of the HQET
formalism for Dirac fermions. Next we have analyzed the Goldstone
theorem in this approximation and identified the Goldstone modes
corresponding to the broken generators as states containing a heavy quark
and a heavy anti-quark. In nature they should correspond to $\bar b b$
and $\bar c c$ mesons, in which the heavy quark mass dependence has
been removed. By calculating the current-current correlator (for the
broken currents) in this theory we show that the dispersion relation
of these Goldstone modes are
consistent with the Goldstone theorem for non-relativistic theories.

In order to connect this theory of the quark and anti-quark in
the large mass limit with the fundamental one we have, firstly,
calculated
the axial-type current-current correlators starting from the fundamental
theory as well. For this purpose, we have identified the currents in
the fundamental theory which correspond to the conserved currents of the
$U(4)$ symmetry at the level of the HQET. We concluded that the
correlator of the axial type of currents studied from the original
theory does not have a smooth limit when m goes to infinity. That shows
that at the level of current correlators the large mass limit
of the fundamental theory are not reproduced by the HQET
\footnote{$^{\dagger}$}{In a relatively different context anomalous
dependences on large masses have been pointed out in [21]}.

Secondly, we have derived the
generating functional of the HQET including sources for all
currents from the generating functional of the fundamental theory
under the standard HQET assumptions [8,13]. The two generating
functionals differ by a non-trivial non-local ratio of determinants.
The determinant of the denominator corresponds
to the heavy quark limit and it is mass independent while the
determinant of the numerator corresponds to the original theory and has
a non-trivial dependence on the mass. With QCD
switched off, both determinants enjoy the same symmetry
properties,
which shows that at the symmetry level and without gluons present the
two theories are equivalent.
In the presence of QCD the calculation of the determinant of the
original theory is plagued with infrared singularities, which signals
that a non-perturbative analysis is needed in order to draw any solid
conclusion.

When full QCD is switched on the singularity structure of
current-current correlators such as \(52) are known
from phenomenology. With the assumption that poles and cuts of light
neutral mesons are suppressed by powers of $1/m$ the first singularity
encountered when rising the energy are the $\eta_{b}$ , $\Upsilon$ ,
$\eta_{c}$ and $ J/\Psi$ poles depending on the currents and heavy
flavors one considers. By quantum numbers any of the $\eta_{b}$ or
$\Upsilon$ ($J/\Psi$ or $\eta_{c}$)
states with the mass of the quark and antiquark subtracted could play
the role of our Goldstone modes. The residual mass (binding
energy) of these states may be considered as due to an explicit breaking
of the U(4) symmetry caused by hard gluons, the effect of which we have
disregarded.

Let us briefly comment on the case of having $N_{f}$ flavors. In that
case formula (1) is still correct, where now $h_{v}$ is taking values
in the flavor space. In the rest frame it is obvious that (1) enjoys a
$U(4N_{f})$ symmetry. One may use the argument after (16) in order to
prove that the $U(4N_{f})$ symmetry holds in any frame. When QCD is
swiched off, the analysis carried out in sect. 4 is not substancially
modified: $U(4N_{f})$ breaks spontaneously down to $U(2N_{f})\otimes
U(2N_{f})$. At the level of HQET, this implies that particles like the
$B_{c}$ meson could also be regarded as Goldstone modes, once the
$m_{b}+m_{c}$ mass dependence is removed. However, if we take into
account the ratio of determinants in \(ex), which appears when making
the precise connection between the fundamental theory and the HQET ,
the implication above desintegrates. When QCD is switched off, the
strong mass dependence of the $\Gamma^{A}$-type current correlators
calculated from the fundamental theory \(52)-\(53) breaks explicitly
any
symmetry relating $\Gamma^{A}$-type currents of different flavors. As
mentioned before, the QCD effects in these correlators are difficult to
estimate in a reliable way, but it would be very surprising that they
conspire to restore the $U(2N_{f})\otimes U(2N_{f})$ symmetry.

 Finally, we would like to emphasize that our analysis is based on a
theory which includes both quark and anti-quark with the same velocity.
The HQET originally was designed to describe hadrons
with a single heavy quark and in the applications studied in the
literature by now has been mainly used to describe either the quark or
the anti-quark sector but not both. The present analysis could
prove useful in describing physical processes which involve
a heavy quark and a heavy anti-quark. Indeed,
the HQET has been recently applied to such physical processes in the
study of the $B_{c}$ meson [22] or $D - \bar D$ and $B-\bar B$ mixing
[12]. The fact that
there is a spontaneously broken $U(4)$ symmetry in HQET describing a
quark and an anti-quark may tell us something interesting about these
processes.

\bigskip

 {\bf Acknowledgements}

We would like to thank N. Andrei, A. Das, B. Grinstein, P.
Haagensen, Yu. Kubyshin
and A. P. Polychronakos for useful discussions and comments.
We are especially indebted to H. Leutwyler and J.M. Pons for
illuminating
discussions at crucial steps of our work. Thanks are also given to J.L.
Goity and J. Taron for the critical reading of the manuscript.

         \bigskip

\centerline{\bf References}

\item{[1]} M.B. Voloshin and M.A. Shifman, {\it Yad. Fiz.} {\bf 45}
(1987) 463 [Sov. J. {\it Nucl. Phys.} {\bf 45} (1987) 292]; H.D.
Politzer and M.B. Wise, {\it Phys. Lett.} {\bf B206}
(1988) 681; and {\it Phys. Lett.} {\bf B208} (1988) 504.

\item{[2]} N. Isgur and M.B. Wise, {\it Phys. Lett.} {\bf B232}
(1989) 113; and {\it Phys. Lett.} {\bf B237} (1990) 527.

\item{[3]} W.E. Caswell and G.P. Lepage {\it Phys. Lett.} {\bf B167}
(1986) 437;
G.P. Lepage and B.A. Thacker, {\it Nucl. Phys.} {\bf B4} (Proc.
Suppl.) (1988) 199.

\item{[4]} E. Eichen and B. Hill, {\it Phys. Lett.} {\bf B234} (1990)
511.

\item{[5]} H. Georgi {\it Phys. Lett.} {\bf B240} (1990) 447.

\item{[6]} B. Grinstein {\it Nucl. Phys.} {\it B339} (1990) 253.

\item{[7]} B. Grinstein, in {\it Proceedings of the Workshop on High
Energy Phenomenology}, Mexico City Mexico, Jul. 1-10, 1991, 161-216
and in {\it Proceedings Intersections between particle and nuclear physics}
Tucson 1991, 112-125.

\item{[8]} T. Mannel, W. Roberts and Z. Ryzak, {\it Nucl. Phys. }
{\bf B363} (1991) 19.

\item{[9]} M.E. Luke, {\it Phys. Lett.} {\bf B252} (1990) 447;
A.F. Falk and B. Grinstein, {\it Phys. Lett.} {\bf B247} (1990) 406;
A. F. Falk, H. Georgi, B. Grinstein and M.B. Wise {\it Nucl. Phys.}
{\bf B343} (1990) 1;
H. Georgi, B. Grinstein and M.B. Wise {\it Phys. Lett.} {\bf B252}
(1990) 456;
A.F. Falk, B. Grinstein and M.E. Luke {\it Nucl. Phys.} {\bf B357}
(1991) 185;
H. Georgi {\it Nucl. Phys.} {\bf B348} (1991) 293;
N. Isgur and M.B. Wise, {\it Nucl. Phys.} {\bf B348} (1991) 276;
T. Mannel, W. Roberts and Z. Ryzak, {\it Phys. Lett.}
{\bf B271} (1991) 421 .

\item{[10]}
A.F. Falk and B. Grinstein, {\it Phys. Lett.} {\bf B249} (1990) 314;
A. F. Falk, H. Georgi, B. Grinstein and M.B. Wise {\it Nucl. Phys.}
{\bf B343} (1990) 1.

\item{[11]} J.L. Goity, {\it Phys. Rev.} {\bf D46} (1992) 3929;
G. Burdman and J.F. Donoghue, {\it Phys. Lett.} {\bf B280} (1992) 287;
B. Grinstein, E. Jenkins, A.V. Manohar, M.J. Savage and M.B. Wise
{\it Nucl. Phys.} {\bf B380} (1992) 369;
R. Casalbuoni, A. Deandrea, N. Di Bartolomeo, R. Gatto, F. Feruglio and
G. Nardulli, {\it Phys. Lett.} {\bf B294} (1992) 106.

\item{[12]} H. Georgi {\it Phys. Lett.} {\bf B297} (1992) 353; W. Kilian
and T. Mannel {\it Phys. Lett.} {\bf B301} (1993) 382.

\item{[13]} J. Soto and R. Tzani {\it Phys. Lett.} {\bf B297} (1992)
358.

\item{[14]} A.F. Falk, M. Neubert and M. Luke {\it Nucl. Phys.}
{\bf B388} (1992) 363.

\item{[15]} J. Goldstone, {\it N. Cim.} {\bf 19} (1961) 154;
J. Goldstone, A. Salam and S. Weinberg {\it Phys. Rev.} {\bf 127} (1962)
965.

\item{[16]} P.W. Anderson, ``Concepts in solids", Sect. 3.D, W.A.
Benjamin Inc. (1964); ``Basic notions in condensed matter physics",
Benjamin/Cummings (1984).

\item{[17]} T.W.B. Kibble, in {\it Proceedings of the 1967 International
Conference on Particles and Fields,} C.R. Hagen et al, eds., John Wiley
and Sons, New York, 1967.

\item{[18]} C. Carone, {\it Phys. Lett.} {\bf B253} (1991) 408.

\item{[19]} S. Coleman, J. Wess and B. Zumino {\it Phys. Rev. } {\bf
D177} (1969) 2239;  S. Coleman, J. Wess and B. Zumino {\it Phys. Rev.}
{\bf D177} (1969) 2247.

\item{[20]} J. Gasser and H. Leutwyler, {\it Ann. Phys. } {\bf(N.Y.)158}
(1984) 142; J. Gasser and H. Leutwyler, {\it Nucl. Phys.}
{\bf B250} (1985) 465.

\item{[21]} P. Ball, H. G. Dosch and M. A. Shifman, `Remarks on the
Heavy Quark Symmetry in the complex plane', HD-THEP-92-50 preprint .

\item{[22]} E. Jenkins, M. Luke, A.V. Manohar and M.J. Savage, {\it
Nucl. Phys.} {\bf B390} (1993) 463.

\vfill\eject
{\bf Figure Captions}
\bigskip
Fig. 1. Modification of the Dirac vacuum in the HQET. By the field
redefinition $ \Psi = e ^{-im \v v \cdot x} h _ v (x)$, the extra time
dependence of the new field is $ e ^{imv ^0 t}$ for the $ h ^+ _ v (x) $
component and $ e ^{-im v ^0 t} $ for the $ h ^- _ v (x) $ component.
The field redefinition, thus, amounts to lowering the positive Dirac sea
energy by an amount $m v ^0$ and raising the negative Dirac sea by the
same amount. Accordingly, the energy levels $mv ^0$ and $ -m v ^0$
of the original theory become zero energy levels in the effective
theory.

Fig. 2. The infinite degeneracy of the vacuum of this HQET, where quarks
and anti-quarks are included. The $ \gamma_ 5$-symmetry mixes these
degenerate states, since the symmetry operator on the state $ |0;0> $
creates the state $ | \vec k , \sigma ; - \vec k , \sigma ^{\prime}>$
without energy cost.

 \vfill \end